\documentclass[aps,prapplied,12pt,superscriptaddress,notitlepage]{revtex4-2} 
\usepackage[utf8]{inputenc}
\usepackage{amsmath, amssymb, bm}      
\usepackage{graphicx}                   
\usepackage{wrapfig}                    
\usepackage{array}                       
\usepackage{booktabs}                   
\usepackage{setspace}                   
\usepackage{placeins}                   
\usepackage{chngcntr}                   
\usepackage{braket}
\UseRawInputEncoding
\usepackage{xr}                         

\usepackage{tikz}                       
\usetikzlibrary{trees}                  
\begin{document}
\title{Magneto-optic perturbation theory for near-complete violation of Kirchhoff's law of thermal emission at low magnetic fields}

\author{Daniel Cui}\affiliation{Department of Materials Science and Engineering, University of California, Los Angeles, Los Angeles, CA 90095, USA}
\author{Aaswath P. Raman}
\email{aaswath@ucla.edu}
\affiliation{Department of Materials Science and Engineering, University of California, Los Angeles, Los Angeles, CA 90095, USA}

\begin{abstract}

Magneto-optic photonic systems can violate Kirchhoff's law of thermal emission by breaking Lorentz reciprocity. We develop a dispersive perturbation theory yielding an analytical expression for magneto-optical resonance frequency shifts in plasmonic semiconductors under applied magnetic fields. This expression shows the shift is governed by the overlap of the mode's optical spin density with the magneto-optical material. We use this expression to design a III-V metasurface that achieves nonreciprocal emissivity contrast of ~0.8 at only 0.1 T, and demonstrate that the theory can explain order-of-magnitude differences in magnetic field sensitivity between different photonic structures.

\end{abstract}
\maketitle 

Controlling the absorption and emission of thermally generated light is of fundamental importance to a broad range of imaging, sensing and energy devices. In typical linear materials thermal emission is subject to detailed balance and Kirchhoff’s law of thermal radiation, which imposes a fundamental constraint originating from Lorentz reciprocity: absorptivity at a particular wavelength and angle must equal its emissivity \cite{kirchhoff_ueber_1860}. Recently, there has been significant interest in nonreciprocal thermal emission in materials that do not obey Lorentz reciprocity \cite{shayegan_direct_2023,shayegan_broadband_2024,zhang_observation_2025,pajovic_nonreciprocal_2025,wang_maximal_2023}. Such control over thermal radiation relaxes the constraint imposed by Kirchhoff’s law and allows for separate control over emissivity and absorptivity. Moreover, the ability to decouple emissivity and absorptivity has shown to be invaluable to improving the performance of energy harvesting devices such as thermophotovoltaics and multijunction solar cells \cite{jafari_ghalekohneh_nonreciprocal_2022, park_reaching_2022, park_nonreciprocal_2022}. 

In order to realize nonreciprocal thermal emission, many kinds of material platforms have been investigated including Weyl semimetals (WSM) \cite{wu_near-complete_2021,kotov_giant_2018,guo_radiative_2020,zhao_axion-field-enabled_2020} and spatio-temporal modulation in metamaterials \cite{hadad_breaking_2016,li_nonreciprocity_2022,ghanekar_nonreciprocal_2023}. In spite of being intrinsically time-reversal symmetry breaking, the synthesis of WSMs has proven to be experimentally challenging which has left such platforms primarily suitable for theoretical explorations. Setups that leverage spatio-temporal modulation have been proposed as well for nonreciprocity, but they typically require high drive frequencies in the THz regime which make them impractical. 

To date, some of the most effective material platforms for breaking Kirchhoff’s law have relied on III-V semiconductor materials such as GaAs, InAs, and InSb \cite{zhang_observation_2025,shayegan_nonreciprocal_2022,shayegan_broadband_2024,shayegan_direct_2023}. Because of their narrow bandgaps and low electron effective masses, many III-V materials have strong magneto-optic responses that have rendered them suitable for demonstrating nonreciprocal thermal emission and absorption. However, most experimental demonstrations utilize designs that require magnetic field biases on the order of 1-3 T which are too high for most applications \cite{zhang_observation_2025,shayegan_nonreciprocal_2022,shayegan_broadband_2024,shayegan_direct_2023}. Numerical demonstrations have shown near complete violation of Kirkchoff’s law at less than 0.5 T in semiconductor metasurfaces \cite{zhu_near-complete_2014,zhao_near-complete_2019}. Theoretically, temporal coupled mode theory (TCMT) has also been applied to better understand the requirements to maximize the nonreciprocal contrast between emissivity and absorptivity \cite{park_violating_2021,zhao_nonreciprocal_2021}. However, such theory is phenomenological and does not properly elucidate the resonance frequency shift that is seen in many of the numerical and experimental demonstrations. Developing better theoretical intuition for this shift in frequency is critical for maximizing nonreciprocal emissivity and absorptivity contrast in nonreciprocal thermal emitter designs at low magnetic field biases.

In this Letter, we develop a closed-form, analytical expression based on dispersive metamaterial perturbation theory for the resonance frequency shift due to application of magnetic field on a plasmonic system. This expression, a function of the optical spin angular momentum density \cite{vernon_decomposition_2024}, is validated for a range of low magnetic field biases at various incident angles. We also show this expression can be used to design a III-V based metasurface that can achieve near-complete violation of Kirchhoff’s law at a remarkably low magnetic field, 0.1 T, suggesting new opportunities to enhance magneto-optic response for nonreciprocal thermal emission 


We begin by first considering a model photonic system composed of a plasmonic magneto optic material. In the absence of magnetic field bias, $\alpha=\epsilon$ regardless of wavelength, angle, and polarization. Once the magnetic field is turned on in the Voigt configuration, the permittivity tensor $\varepsilon$ acquires off-diagonal components

\begin{align}
    \begin{split}
       \varepsilon &= \begin{pmatrix} \varepsilon_{xx} & i\varepsilon_{g} & 0 \\
       -i\varepsilon_{g}  &  \varepsilon_{xx} & 0 \\
       0 & 0 & \varepsilon_{zz} \\
       \end{pmatrix}
    \end{split}
    \label{eq:perm_tensor}
\end{align}

A non-zero magnetic field results in non-zero off-diagonal components that cause the system to break Lorentz reciprocity, and in turn implies that the absorptivity and emissivity are no longer equal  $\alpha \neq \varepsilon$. In mirror symmetric structures there is a stronger constraint on the relationship between absorptivity and emissivity, $\alpha(\lambda, \theta,\hat{p}) = \varepsilon(\lambda, -\theta,\hat{p})$\cite{guo_theoretical_2020,zhao_nonreciprocal_2021}. The overarching goal of research in this topic has been to  maximize the nonreciprocal contrast $\eta = |\alpha - \varepsilon|$ at a particular angle and wavelength. Here, we seek to derive a set of conditions that can guide the photonic design of a structure to maximize this contrast and thereby maximally violate Kirchhoff's law. 

To that end, temporal coupled mode theory (TCMT) can provide useful insight due to the coupling of guided resonances with multiple free space diffraction channels \cite{zhao_nonreciprocal_2021,park_violating_2021,miller_universal_2017}. To maximize the absolute difference between emissivity and absorptivity in a certain channel, we seek a photonic structure that achieves the critical coupling condition $\gamma_i = \gamma_r$ where the intrinsic loss and radiative loss rates equal each other, and to maximize the difference between the in-coupling and out-coupling rates\cite{park_violating_2021,fan_theoretical_1999}. However the dominant effect underlying Kirchhoff's law violation is the frequency shift of a mode due to the applied magnetic field. Current approaches do not provide either a prediction for this frequency shift or an analytical understanding of the underlying mode characteristics that optimize this behavior.  Motivated by this gap, we develop a perturbation theory for plasmonic semiconductors by treating the magnetic field bias as a small perturbation.

The dielectric permittivities in Eq. (1) obeys a Lorentz-Drude model \cite{raman_perturbation_2011,raman_photonic_2010,joannopoulos2008molding}. In the presence of an external magnetic field, the free carriers experience both the electric field and the magnetic Lorentz force.  In the Voigt configuration as indicated in Eq.~(\ref{eq:perm_tensor}) the on-diagonal and off-diagonal permittivity terms adopt the following frequency-dependent Drude-Lorentz expression \cite{shayegan_direct_2023}:

\begin{align}
    \begin{split}
\varepsilon_{xx}&=\varepsilon_{yy} = \varepsilon_{\infty}-\frac{\omega^2_p(\omega+i\Gamma)}{\omega[(\omega+i\Gamma)^2-\omega^2_c]} \\
\varepsilon_{zz} &= \epsilon_{\infty} - \frac{\omega^2_p}{\omega(\omega+i\Gamma)}\\
\varepsilon_{g} &=\frac{\omega^2_p\omega_c}{\omega[(\omega+i\Gamma)^2-\omega^2_c]}
    \end{split}
    \label{eq:V_eom}
\end{align}
Here, $\omega_p$ and $\omega_c$  are the plasma and cyclotron frequencies respectively, while $\epsilon_{\infty}$ and $\Gamma$ are the background dielectric constant in the high frequency limit and scattering rate respectively. Such models of the permittivity start from the classical damped oscillator equation of motion subject to the full Lorentz force, and accurately capture the optical response of semiconductor materials over infrared wavelengths. 

Defining the macroscopic polarization $\overrightarrow{P} = Ne\overrightarrow{r}$ together with the auxiliary polarization velocity field $\overrightarrow{V} = d\overrightarrow{P}/dt$ \cite{joseph_direct_1991}, the material response can be expressed as a first-order equation of motion:
\begin{align}
    \begin{split}
        \frac{d \overrightarrow{V}}{dt} + \Gamma\overrightarrow{V} + \omega^2_0\overrightarrow{P} = \omega^2_p \varepsilon_{\infty} \overrightarrow{E} + \omega_c \overrightarrow{M} \times \overrightarrow{V}
    \end{split}
    \label{eq:V_eom}
\end{align}
where $\omega_p$ is the plasma frequency, $\omega_0$ is the resonance frequency, $\Gamma$ is the scattering rate, and $\omega_c = eB/m$ is the cyclotron frequency that encodes the applied magnetic field strength. The unit vector $\overrightarrow{M}$ denotes the direction of the applied magnetic field. The key observation is that the applied magnetic field enters Eq.~(\ref{eq:V_eom}) solely through the cross-product term $\omega_c \overrightarrow{M} \times \overrightarrow{V}$, which couples the orthogonal components of the polarization velocity and is directly proportional to the applied field strength.

Combined with Maxwell's curl equations for $\overrightarrow{E}$ and $\overrightarrow{H}$, Eq.~(\ref{eq:V_eom}) yields a closed system of first-order partial differential equations for the full state vector $\mathbf{x} = (\overrightarrow{H},\, \overrightarrow{E},\, \overrightarrow{P},\, \overrightarrow{V})^T$. For plane-wave fields, this system can be cast into a generalized eigenvalue problem of the form $\omega A \mathbf{x} = B \mathbf{x}$ \cite{raman_perturbation_2011,raman_photonic_2010}, where $A = \text{diag}(\mu_0,\, \varepsilon_{\infty},\, \omega^2_0/\omega^2_p\varepsilon_{\infty},\, 1/\omega^2_p \varepsilon_{\infty})$ is a positive-definite diagonal operator encoding the energy density of the dispersive system, and $B$ encodes the spatial and temporal coupling between the field components (see Supplemental Material for the explicit matrix forms). Because the $\omega_c$ term in Eq.~(\ref{eq:V_eom}) acts only on $\overrightarrow{V}$, the applied magnetic field modifies only the operator $B$ as $B \to B + D$, where $D = \text{diag}(0,\, 0,\, 0,\, i\omega_c \overrightarrow{M}\times / \omega^2_p \varepsilon_{\infty})$.
For the lossless case ($\Gamma = 0$) with TM-polarized radiation ($E_z = 0$), first-order perturbation theory of the generalized eigenvalue problem yields (see Supplementary Information):
\begin{align}
    \Delta \omega = \frac{\braket{\mathbf{x}|D|\mathbf{x}}}{\braket{\mathbf{x} | A | \mathbf{x}}}
    \label{eq:pert_formula}
\end{align}
where $\mathbf{x}$ is the unperturbed eigenstate. The denominator $\braket{\mathbf{x}|A|\mathbf{x}}$ is the total electromagnetic energy density $W$ of the mode, which includes both the standard field energy and the kinetic energy stored in the dispersive polarization currents:
\begin{align}
    \begin{split}
       W = \iint \frac{1}{2}\left[\varepsilon_{\infty} (|E_x|^2 + |E_y|^2) + \mu_0 |H_z|^2\right] + \frac{1}{2\varepsilon_{\infty} \omega^2_p} \left[|V_x|^2 + |V_y|^2 + \omega^2_0 (|P_x|^2 + |P_y|^2)\right] dx\, dz
    \end{split}
    \label{eq:energy_density}
\end{align}
where we assume translational invariance in the $y$ direction. Note that the last two terms, arising from the auxiliary fields, capture the dispersive correction to the energy density that is absent in non-dispersive materials. Evaluating the numerator $\braket{\mathbf{x}|D|\mathbf{x}}$ for a Drude material ($\omega_0 = 0$), we can relate the auxiliary polarization velocity to the electric field via $\overrightarrow{V} = -i\omega^2_p\varepsilon_{\infty}\overrightarrow{E}/\omega$ at the unperturbed eigenfrequency. Substituting this into the numerator and combining with Eq.~(\ref{eq:energy_density}), we obtain our central result:
\begin{align}
    \begin{split}
        \Delta \omega = \frac{ \frac{-\omega_c\omega^2_p\varepsilon_{\infty}}{\omega^2}   \iint \text{Im}(E^*_x E_y -E^*_y E_x )\, dx\, dz}{\iint \frac{1}{2}[\varepsilon_{\infty} (|E_x|^2 + |E_y|^2) + \mu_0 |H_z|^2] + \frac{1}{2\varepsilon_{\infty} \omega^2_p} (|V_x|^2 + |V_y|^2)\, dx\, dz}
       \end{split}
       \label{eq:delta_omega}
\end{align}
where all fields are evaluated at the unperturbed ($B = 0$) eigenfrequency $\omega$ and the integration in the numerator extends over the magneto-optical region.
Equation~(\ref{eq:delta_omega}) provides an intuitive physical picture for the frequency shift due to the applied magentic field. The frequency shift is proportional to the overlap integral of the $z$-component of the optical spin angular momentum density, $s_z \propto \text{Im}(\overrightarrow{E}^{*} \times \overrightarrow{E})_z = \text{Im}(E^*_x E_y - E^*_y E_x)$, with the magneto-optical material \cite{vernon_decomposition_2024}. This quantity measures the local degree of circular polarization of the mode. It is zero for linearly polarized fields (where $E_x$ and $E_y$ are in phase) and is maximized when the field components are $\pi/2$ out of phase. This favors structures where the magneto-optical layer directly participates in supporting the guided resonance, rather than acting as a thin perturbative film. The denominator is the total electromagnetic energy density of the mode, including the kinetic energy stored in the dispersive polarization currents. 

Consequently, Eq.~(\ref{eq:delta_omega}) suggests that mode profiles should be engineered to produce strong local elliptical polarization within the magneto-optical layer and maximize the spatial confinement and overlap of the mode with the magneto-optical material. Additionally, since $\omega_c = eB/m$ and $\omega^2_p = Ne^2 / m\varepsilon_0$, the material-dependent prefactor in the numerator scales as $\omega_c \omega^2_p \propto NB/m^2$. This highlights why low effective mass $m$ is one of the most important material parameters, entering as it does quadratically. Indeed, InAs and InSb, with some of the lowest effective masses accessible in III-V semiconductors ($m \approx 0.033\,m_e$ and $m \approx 0.014\,m_e$ respectively), have performed well in both numerical and experimental demonstrations of nonreciprocal thermal emission. 

To validate our theoretical results we consider magneto-optic InAs layers lying beneath a one-dimensional grating composed of undoped InGaAs whose grating width $d = 5 mu$m and periodicity $\Lambda = 10 \mu$m. The height of the grating is $t_1 = 0.27 \mu$m which sits on top of an undoped InGaAs substrate of $t_2 = 4 \mu$m. This lossless dielectric layers sits on top of a $t_3 = 5 \mu$m thick n-type InAs layer with doping $7.8 \times 10^{17}$ cm$^{-3}$, $\Gamma = 1.55 \times 10^{11} s^{-1}$ and $m = 0.033$ $m_e$. Lastly, a $5e19 \; cm^{-3}$ InAs layer with $\Gamma = 1e13 \; 1/s$ with $m = 0.15 \; m_e$ is optically thick and acts as a perfect reflector at the bottom of the structure. The plasma frequencies used in Equation (2)-(4) are $\omega_p = \sqrt{\frac{ne^2}{m\varepsilon_0}}$ and $\varepsilon_{\infty} = 12.3$.  The material parameters values were obtained using a semiconductor model previously used to model nonreciprocal thermal emitters\cite{zhu_near-complete_2014}. For InGaAs, we use the refractive index model given in Ref. \cite{le_determination_2022} and note that it behaves as a high-index lossless dielectric coupling layer.

The resulting structure supports a high Q resonance at frequencies between 12.2210 and 12.310 THz.  When a magnetic field bias of +0.1 T is turned on, the resonance frequency redshifts to lower frequencies (Figure. 1(b)). The parameters of the geometry were also tuned to ensure the critical coupling condition $\gamma_i=\gamma_e$ was satisfied as shown by the near-maximized unity emissivity peak at 0 T  (Figure 2(a)). Due to the combination of high Q factor, critical coupling, and substantial frequency shift, we observe minimal overlap between the 0 T and 0.1 T resonance peaks for incident angles between 67 and 80 degrees, yielding near-complete violation and nonreciprocal contrast  $\varepsilon_{0.1T} - \varepsilon_{0T}$ between 0.7 and 0.8. 

\begin{figure}[ht]
    \centering

    \includegraphics[width = 5.5in]{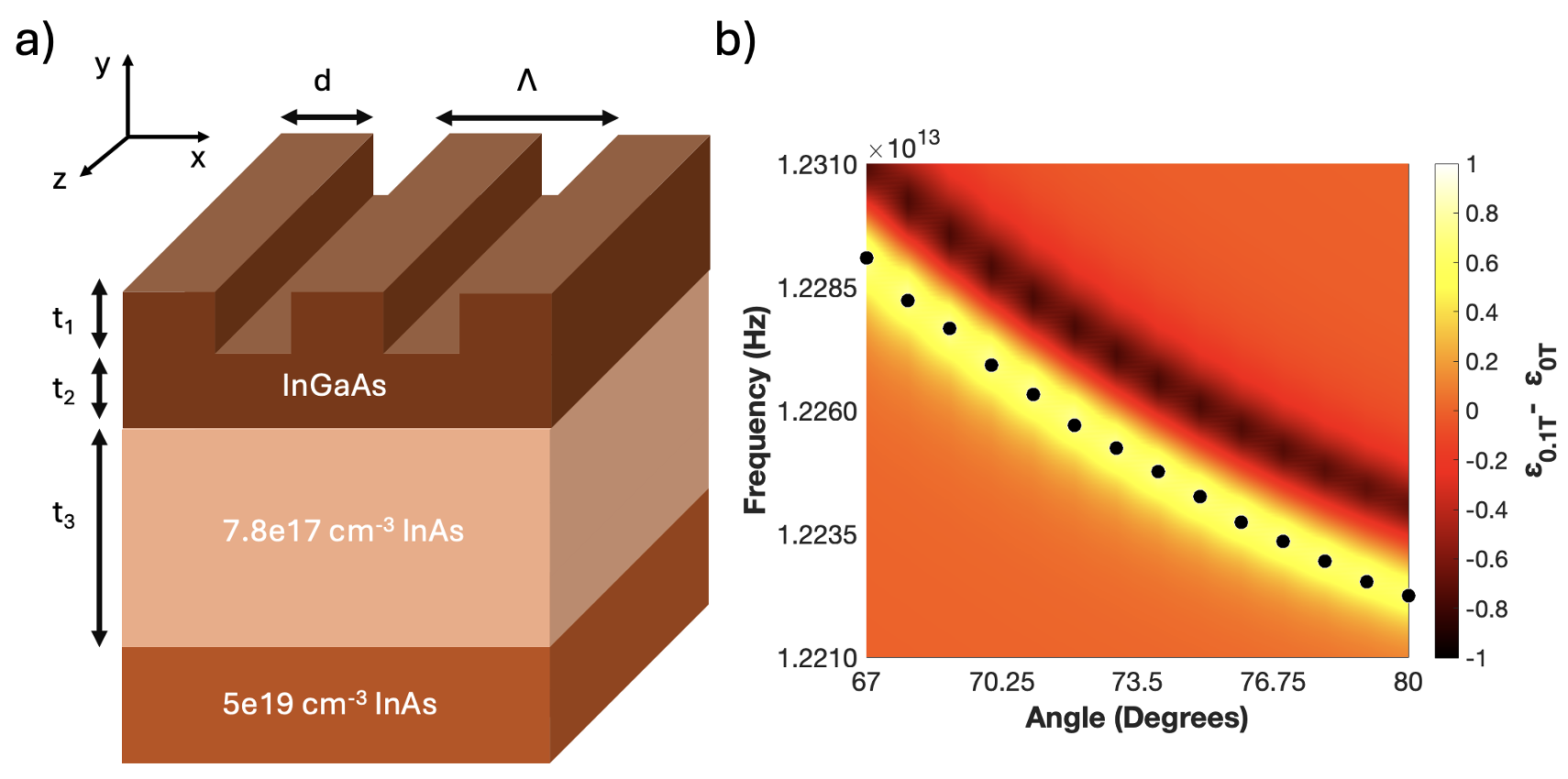}
    \caption{(a) Schematic of the structure used for theory verification. From top to bottom, the structure consists of a undoped InGaAs grating with d = 5 $\mu$m and grating period $\Lambda$ = 10 $\mu$m that is 0.27 $\mu$m tall, undoped InGaAs substrate that is 4 $\mu$m thick, lightly doped InAs that is 5 $\mu$m thick, and a heavily doped InAs layer that is optically thick enough to act as a perfect reflector. (b) Resulting color plot that shows $\varepsilon_{0.1T}-\varepsilon_{0T}$ for a p-polarized incident wave. An external magnetic field is applied in the z direction within both the InAs layers. The bright yellow band shows the shifted resonance peaks upon application of magnetic field while the darker band are the original locations of the 0 T peaks.The black dots are the perturbation theory predictions at a particular incident angle.}
    \label{fig:wavel_vs_angle}
\end{figure}

As a demonstration of our perturbation theory, we use the expression derived in equation 15 to calculate the frequency shifts at 0.1 T. The effect of the off-diagonal permittivity terms are negligible in the higher doped InAs reflector as the $\frac{|\varepsilon_{xy}|}{|\varepsilon_{xx}|}$ is around 0.0016 within the target frequency range due to the high values of the on-diagonal terms. In the lower doped InAs layer, $\frac{|\varepsilon_{xy}|}{|\varepsilon_{xx}|}$ is about 0.26, suggesting that the magneto optical effects are multiple orders of magnitude stronger in the top InAs layer. Thus, we note that all parameter and field values used for the first order frequency shifts were taken at the unperturbed 0 T case from the lightly doped InAs layer. Furthermore, for the sake of elucidating the first order frequency shift in $Re(\omega_0)$ due to the magneto optical effect, we ignore the effect on $Im(\omega)$ from the on-diagonal and off-diagonal terms of the permittivity tensor. As $Im(\omega)$ controls the linewidth, this choice is justified by the fact that the magnetic field perturbation does not substantially change the full width half maximum of the resonance. 

\begin{figure}[ht]
    \centering
    \includegraphics[width = 6in]{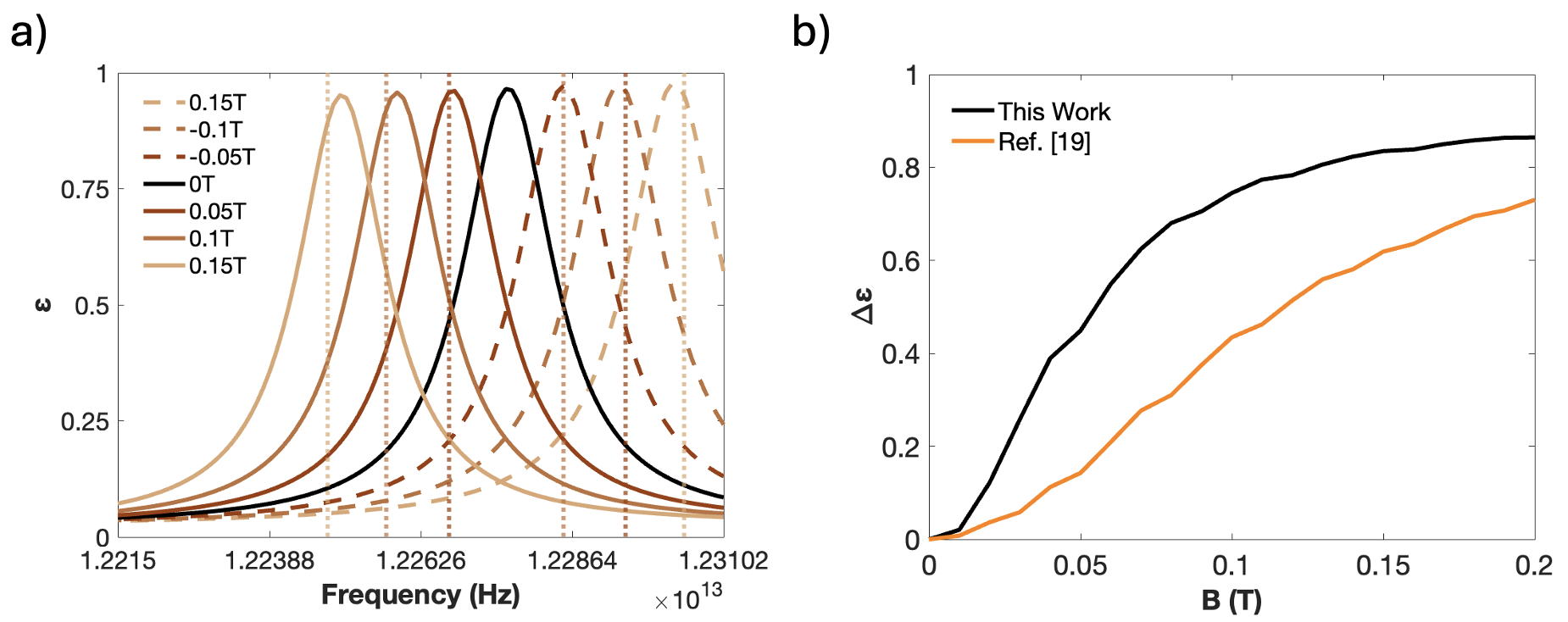}
    \caption{(a) Emissivity spectra for a range of magnetic fields between 0.15T and -0.15T at 0.05T intervals. The vertical lines are the resonance frequency shifts predicted by the pertubation theory from the 0 T resonance frequency. (b) the change in emissivity where $\Delta \varepsilon = \varepsilon_{0.1T} - \varepsilon_{0T}$ as the applied magnetic field is swept from 0 T to 0.2T. To provide a sense of the high magnetic field sensitivity, the structure in reference [19] is used as a comparison.} 
    \label{fig:B_field}
\end{figure}

As can be seen in Figure 1(b), there is significant nonreciprocal contrast (~0.8) upon applying a very modest 0.1 T magnetic field. The dark band represents the location of the 0 T emissivity resonance peaks while the brighter band are the shifted emissivity peak locations upon application of a 0.1 T magnetic field. Thus, upon application of our perturbation theory, we can see in Figure 1(b) that shifted resonance frequency locations, indicated by black dots, that were predicted by the perturbation theory lie within the center of the bright band of resonances. The theoretical prediction matches very well with the 0.1 T resonance peak locations across all angles of incidence.

We also examine the validity of the perturbation theory with varying mangetic field strengths. We focus in particular on a polar angle of 72 degrees at which the structure exhibits optimal nonrecriprocal contrast. As shown in Figure 2(0a, we plot the resulting spectra by sweeping from -0.15 T to 0.15 T in increments of 0.05 T.  Negative magnetic fields are associated with a blue shift of the resonance frequency while positive magnetic fields result in a redshift. Remarkably, we note that the difference $\varepsilon_{B\neq0~T} - \varepsilon_{0~T}$ is above ~0.9 at a modest applied magnetic field of 0.15 T. The vertical lines plotted in Figure 2(a) are the calculated shifted frequencies using the perturbation theory of Equation (\eqref{eq:pert_formula}). There is excellent agreement between the predicted frequencies at nonzero magnetic field versus the directly simulated peak locations. However, as the magnitude of the magnetic field perturbation is increased up to 0.2 T the perturbation theory's prediction begin to deviate from the simulated emissivity peaks. This is in line with the basic assumptions of the theory that the magnetic field perturbation is small and is a first-order effect. 

In the context of nonreciprocal thermal emission control, this particular design and structure offers exceptional performance at low magnetic fields, a practically important outcome. The structure is able to achieve a strong nonreciprocal contrast ~0.5 for a 0.05 T magnetic field. Such field strengths can be easily accessed with simple permanent magnets. The high magnetic field sensitivity is illustrated in Figure 2(b) where the change in emissivity at the resonance frequency is plotted against the magnetic field strength. The initial slope between 0 and 0.05T indicates that the slope $\Delta B/\Delta \varepsilon$ is ~ 10. In comparison, the GMR structure used in Ref. [19] has a high but more modest sensitivity of ~3  within the same magnetic field range. 

 \begin{figure}[ht]
    \centering
    \includegraphics[width = 6.5in]{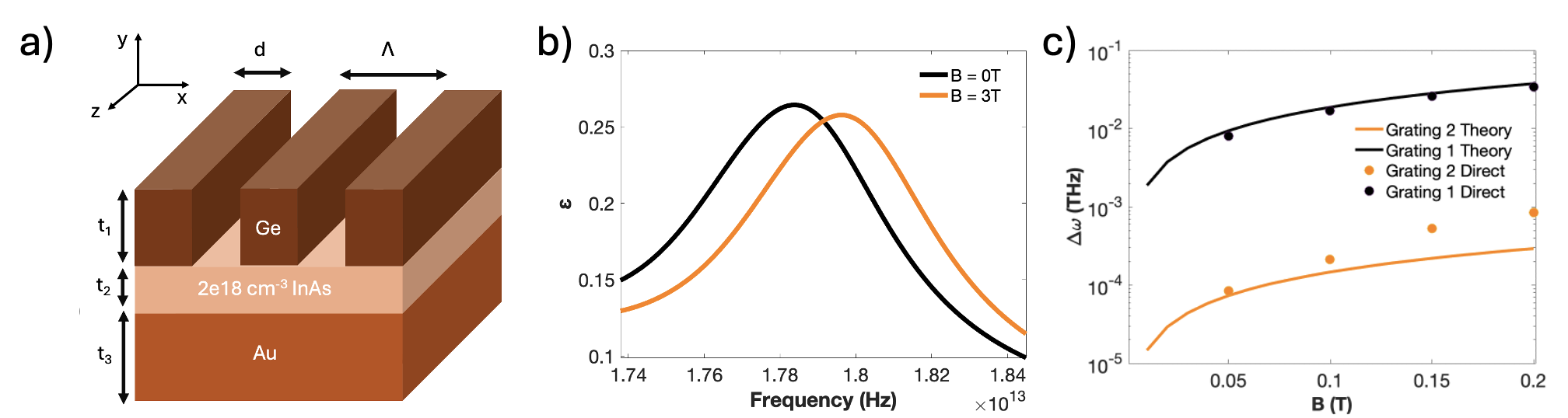}
    \caption{(a) Schematic of a second GMR structure  with Germanium on top of a highly doped InAs layer on top of a Au reflector layer. (b) Emissivity spectra at 77 degrees plotted at 0 T and 3T. (c)Comparison of the $\Delta \omega$ frequency shift values as the applied magnetic field strength is increased. The solid lines indicate the perturbation theory prediction for grating structures 1 and 2, while the dots are the frequency shifts calculated directly from FEM simulations.}
    \label{fig:worse_structure}
\end{figure}

To elucidate the origins of the exceptional strength of Kirchhoff's Law violation in structure 1 in Figure 1(a), we contrast this high-performing structure against a grating structure that has far worse magnetic field sensitivity and nonreciprocal contrast. As shown in Figure 3(a), this second structure consists of a Germanium grating on top of a highly doped InAs layer and a gold substrate\cite{olmon_gold_2012}. The material parameters for InAs are $\omega_p 2 \times 10^{18}$ cm$^{-3}$, $\Gamma = 3.33 \times 10^{12}$ s$^{-1}$, and $m_{eff} = 0.041*m_e$. The Germanium grating, whose refractive index is given by Ref. \cite{li_germanium_1980}, has a width of 2 $\mu$m and period of 9 $\mu$m with a height of 500 nm. The  InAs layer is extremely thin at 10 nm. Finite element simulations are used to calculate the emissivity spectra at 0 T and 3 T at a polar angle of incidence of 77 degrees (Figure 3(b)). When comparing the 0 T spectrum in Figure 3(b) to the spectrum in Figure 1(b), it is evident that the emissivity peak is much broader in structure 2 than in structure 1 owing to the higher loss of the InAs at these frequencies ($\varepsilon_{xx}=\varepsilon_{yy}=\varepsilon_{zz} \approx -0.04+0.3667i$). This significantly decreases the Q-factor of the resonance and necessitates much higher magnetic field strengths to realize noticeable nonreciprocal contrast. It is also important to note that structure 2's resonance in Figure 3(b) is tuned farther away from critical coupling with significant reflection than structure 1 due to weak coupling of the external radiation to the guided resonance.
\begin{figure}[ht]
    \centering
    \includegraphics[width = 6in]{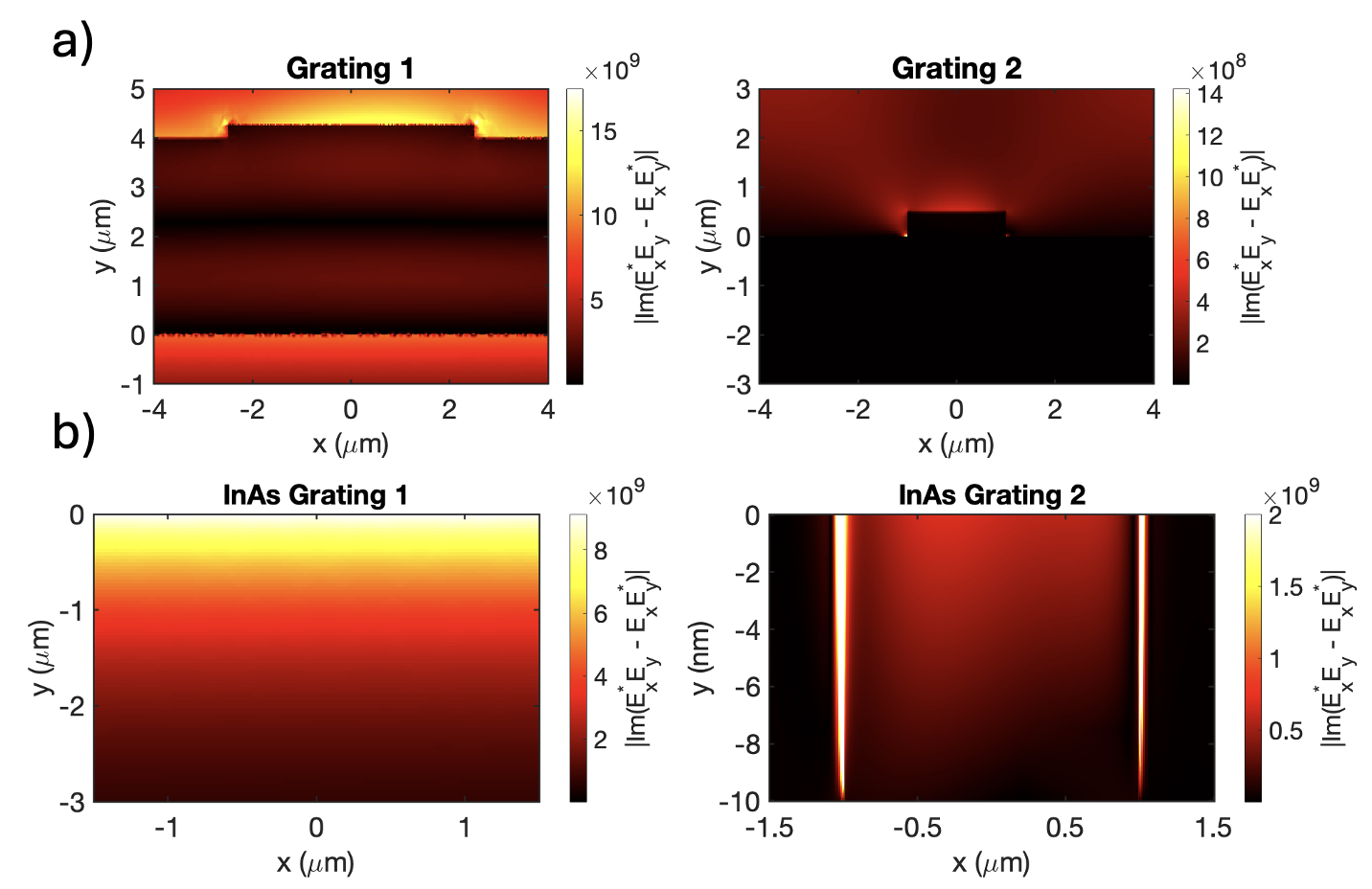}
    \caption{(a) Comparisons of the field expression $|Im(E^*_x E_y- E_x E^*_y)|$ between the 1st grating at $72^{\circ}$ and 2nd grating at $77^{\circ}$.  (b) Comparisons of the field expression $|Im(E^*_x E_y- E_x E^*_y)|$ within the InAs layer between structure 1 at $72^{\circ}$ and structure 2 at $77^{\circ}$.}
    \label{fig:field_plots}
\end{figure}

Our key observation here is that structure 2's resonance is much less sensitive to the applied magnetic field. The significant observed difference in magnetic field sensitivity between the two structures can be attributed to a few major factors based on the developed perturbation theory (Figure 3(c)). As with structure 1, the frequency shifts that are calculated by the perturbation theory and directly from FEM simulations in structure 2 show good agreement at low magnetic field strengths less than 0.1 T. At higher magnetic field strengths, the directly calculated shifts from FEM begin to deviate more prominently than in structure 1 from the first-order shifts predicted by theory. Nevertheless, there is a clear 2 orders of magnitude difference between the resonance frequency shifts in structures 1 and 2 given the same magnetic field strength.

This can be understood by visualizing the optical spin angular momentum density in Figure 4. The spin angular momentum density is significantly higher in intensity in structure 1. Additionally the mode is more strongly confined throughout the magneto-optical InAs layer in structure 1 relative to structure 2. This is since structure 2 is simply too thin to support a guided mode on its own, with the mode overlap instead being primarily in the non-magneto-optic germanium layer. The higher effective mass in structure 2 also reduces the cyclotron frequency $\omega_c = eB/m$ by roughly 20\% for the same applied field, further diminishing $|\Delta\omega|$. 

In conclusion, we have developed a perturbation theory analysis of the magneto optic responses of plasmonic semiconductor materials. We show that this theory is accurate and validated for a range of incident angles and magnetic field strengths for a photonic structure that shows nonreciprocal contrast of ~0.8 at 0.1 T. More importantly, this better elucidates the mechanisms for near-complete violation of Kirchhoff's law at the smallest of magnetic fields possible by showing that the mode profile of the p-polarized incident radiation needs to be optimally engineered to overlap with the magneto optical material. Our work opens the door to new theoretically-grounded strategies for realizing greater nonreciprocal contrast between emissivity and absorptivity at lower, more practical magnetic field strengths. This in turn may make nonreciprocal thermal emission more accessible from a technological point of view.

\section{Acknowledgments}
This material is based upon work supported by the National Science Foundation under grants No. ECCS-2146577. D. C. was supported by the National Defense Science and Engineering Graduate (NDSEG) Fellowship. 

\bibliography{Ref}

\counterwithin{figure}{section}

\clearpage

\end{document}